\documentclass[conference]{IEEEtran}
\usepackage{cite}
\ifCLASSINFOpdf
   \usepackage[pdftex]{graphicx}
\else
\fi
\usepackage{array}
\usepackage{eqparbox}
\usepackage[tight,footnotesize]{subfigure}
\usepackage{url}
\usepackage{color}

\begin{document}
\title{Using Open Standards for Interoperability\\
{\huge Issues, Solutions, and Challenges facing Cloud Computing}}

%\author{\IEEEauthorblockN{Piyush Harsh}
%\IEEEauthorblockA{Institut national de recherche\\ en informatique et en
%automatique\\ INRIA - Team Myriads\\
%Rennes, Bretagne 35042 France\\
%Email: http://info-piyush.rhcloud.com/}
%\and
%\IEEEauthorblockN{Yvon Jegou}
%\IEEEauthorblockA{INRIA - Team Myriads\\
%Rennes, Bretagne 35042 France\\
%Email: yvon.jegou@inria.fr}
%\and
%\IEEEauthorblockN{Roberto Cascella}
%\IEEEauthorblockA{INRIA - Team Myriads\\
%Rennes, Bretagne 35042 France\\
%Email: roberto.cascella@inria.fr}
%\and
%\IEEEauthorblockN{Christine Morin}
%\IEEEauthorblockA{INRIA - Team Myriads\\
%Rennes, Bretagne 35042 France\\
%Email: christine.morin@inria.fr}}


\author{\IEEEauthorblockN{Piyush Harsh\IEEEauthorrefmark{1},
Florian Dudouet\IEEEauthorrefmark{1},
Roberto G. Cascella\IEEEauthorrefmark{1},
Yvon Jegou\IEEEauthorrefmark{1},
 and
Christine Morin\IEEEauthorrefmark{1}}
\IEEEauthorblockA{\IEEEauthorrefmark{1}Inria Rennes - Bretagne Atlantique\\
France\\
Email: \{piyush.harsh, florian.dudouet, roberto.cascella, yvon.jegou,  
christine.morin\}@inria.fr}
}


\maketitle

\begin{abstract}
%\boldmath
Virtualization offers several benefits for optimal resource utilization over
traditional non-virtualized server farms. With improvements in internetworking
technologies and increase in network bandwidth speeds, a new era of computing
has been ushered in, that of grids and clouds. With several commercial cloud
providers coming up, each with their own APIs, application description formats,
and varying support for SLAs, vendor lock-in has become a serious issue for
end users. This article attempts to describe the problem, issues, possible
solutions and challenges in achieving cloud interoperability. These issues will
be analyzed in the ambit of the European project Contrail that is trying to adopt
open standards with available virtualization solutions to enhance users' trust
in the clouds by attempting to prevent vendor lock-ins, supporting and
enforcing SLAs together with adequate data protection for sensitive data.
\end{abstract}


% \fi
\IEEEpeerreviewmaketitle

\section{Introduction}
With the improvements in network bandwidth, more interesting uses of the
Internet have emerged in recent times. One such use is remote execution of
tasks on distant physical machines. The tremendous body of work in grid
computing paved the way for better commercial utilization of the technology for
general purpose computing tasks such as web hosting, data aggregators using
map-reduce, scientific and commercial work loads within the ambit of cloud
computing.

Cloud computing has come as a blessing for small-mid-scale enterprises, SMEs.It
has allowed companies to lease computing infrastructures at economical rates
and has reduced the infrastructure entry barrier for new companies
significantly. As more and more uses of cloud computing are being explored,
the technology faces new challenges that need timely intervention for the
pace of adoption of cloud computing to be sustained and even improved.

Many traditional software and services companies have already jumped on the
cloud computing bandwagon. Notable among them are public Cloud offerings from
Amazon~\cite{AWS}, Google~\cite{AppEngine}, and Microsoft~\cite{Azure}. In
addition service bigwigs such as IBM~\cite{IBMCloud} and HP~\cite{HPCloud} are
not far behind either. There is tremendous activity in improving the hypervisor
technology going on in both commercial as well as in the realm of open source
software development. A hypervisor is a hardware virtualization software that
allows multiple operating systems to run on the same physical machine. The term
\textit{hypervisor} is in a sense superlative of the term \textit{supervisor}.
It is the central element in any cloud computing offering. With the multitude of
virtualization solution available from both commercial vendors such as VMWare
and Citrix, as well as reasonably mature open source free technologies such as
KVM~\cite{KVM}, XEN~\cite{XEN}, VirtualBox~\cite{VirtualBox}, etc, the pace of
research exploring value addition on top of such technologies has picked up
tremendously in recent years.

But with the ease of movement of computation and data in the clouds, comes
numerous challenges that must be addressed promptly. Some of the challenges
belong in the realm of traditional network security research, but many more new
challenges are coming up, some that are beyond the realm of computer science to
solve and require a new and radical scrutiny of international data and privacy
laws and legal jurisdiction in this interconnected world that is no longer bound
by physical boundaries.
%\comment{It is not clear what is the relationship of this
%last sentence. How do you fit this in the global picture?}
%\comment{\color{red}{Well I just wrote that there are numerous issues and just
%for the sake of completeness I also included the legal challenges even though
%it
%is outside the scope of our research to solve.}}

This paper will try to touch upon few of these issues. It will try to explore
possible solutions and analyze the challenges in the remaining issues which are
bound to plague the world of cloud computing sooner or later. Cloud
interoperability has emerged as a major issue. The prospect of
vendor lock-ins may be keeping big customers including governments, healthcare,
and banking away from the clouds. Hence addressing the issue of
interoperability and portability is both timely and necessary.

In order to better define the problem of interoperability, it is important to
understand what is interoperability. Typically it means the ability of different
heterogenous systems to be able to function/interact together. For clouds,
interoperability could be defined as the ability to understand each others
application formats, Service Level Agreement (SLA) templates, authentication and
authorization token formats and attribute data. Although this paper will
identify several challenges in this field, the main focus will be limited to
investigating the problem of cloud interoperability.

\section{Potential and Challenges}
With cloud computing, a person can lease a large number of compute nodes made
available by an Infrastructure as a Service (IaaS) provider. The user can then
create virtual machines configured to run the desired application and deploy
them over the leased compute nodes. Another scenario, a user could request the
services from Platform as a Service (PaaS) provider that already hosts services
required by the end user to execute her application. It could be a simple PHP
code requiring the services of a simple SQL database. In this scenario, the
user just needs to focus her energy on service development and not worry about
where and how to setup the virtual machine to host and execute her services. In
a more complex example, the user can combine services from different kind of
providers, use a PaaS provider in conjunction with a Storage as a Service
(SaaS) provider to lease a chunk of online data space to hold the data, log
files, configurations, etc. There are even attempts to provide Service as a
Service to the end user where they can use service composition to create a
more complex service. The scope of innovation with cloud computing seems
limitless today.

For big enterprises, the biggest asset is the intellectual property and the
knowledge they have in the data they own. They are rightly reluctant in putting
and hosting such data in an environment where they do not maintain absolute
control.

Another constraint comes from the data protection and privacy laws enforced by
different governments. Such laws call for strict geo-location restriction of
data hosting and movement.

Then there comes the problem of SLAs between the end
users and the provider. How does one verify that the agreed SLAs were honored in
the first place, and if violated how can such violations be proved for possible
claims and settlements?

Another concern is disparity in cloud APIs provided by different vendors to the
end users. Such disparity results in vendor lock-in situations where a user is
unable to migrate her cloud deployment over to another cloud provider because
of interface incompatibilities between the two.

There can be security issues arising from colocation of multiple applications in
the same physical local area network (LAN). How does one enforce traffic
isolation between different applications. How does one provide VM protection
from a malicious cloud application being hosted inside the same cluster and LAN?

While there are many more challenges that needs addressing, the one we will
focus in this article is achieving interoperability between multiple cloud
providers. The interoperability is an important aspect from the perspective of
an end user which to some extent addresses the problem of vendor lock-ins. We
will point out challenges to true interoperability and present the current
landscape of the community and industry efforts in this direction.
%\comment{It is really important to define what is meant as interoperability and
%at which level we can achieve it. In my opinion there is no clear flow of
%information between the introduction, potential challenges and then the rest.
%We
%must have a well identified problem statement, that somehow has been given as
%granted.}

\section{Essential elements of a distributed cloud application}
Before beginning to address the problem of interoperability, it is important to
first understand the key components making up a typical cloud application. It
makes sense to look at IaaS services to get the true picture as other forms of
cloud services such as PaaS and SaaS are value addition on top of IaaS services.

A cloud application to be hosted over IaaS clouds comprise of sets of VMs
possibly linked with each other in a private LAN with access to/from external
Internet through a gateway or a proxy. Therefore the critical elements of an
IaaS cloud application are:
\begin{itemize}
  \item Virtual Machines description;
  \item Virtual Network elements linking VMs;
  \item OS image files to run inside the VMs;
  \item Data Stores to be attached to the VMs.
\end{itemize}

Apart from the bare-minimum requirements that has been listed above, the user
would also require some formalism in the agreement between herself and the
provider. These would include such elements as:
\begin{itemize}
  \item Service Level Agreements (SLAs);
  \item Placement restrictions and data protection agreements (QoP);
  \item Billing and auditing provisions for compliance tests and verification;
  \item Monitoring mechanisms for the user to infer the health of the deployed
application.
\end{itemize}

With the pieces in the puzzle identified, one can start to look into how to
achieve interoperability between different cloud providers. For full
interoperability, each of the above identified piece must be easily portable
between different providers, at least in the format and the processes involved.

\section{Standards Landscape}
Open standards are the main proponents of interoperability. An inclusive
standardization process has more milage in getting accepted by the stakeholders
than a process that is exclusive. A question may be raised regarding the
adoption of cloud standards by well entrenched providers. Why is being
interoperable good for them? Critics always point that being able to
vendor-lock-in a customer is good for the business as it may reduce customer
churn, but in our opinion this may not be true. Brand loyalty can be achieved
by providing superior services at attractive prices. Further being
interoperable could bring in big government, banks, and health-care providers'
businesses into clouds thus vastly increasing the customer base.

%\comment{What is the reason for
%getting solutions adopting standards acceptable by stakeholders? Why
%stakeholders? If a stakeholder invests would like to be sure that he has a
%certain return from the business, and a close solution should be more
%profitable
%for him. I would say more for business customers, because they can port their
%applications over different platforms implementing standards. This opens a
%competitive market that will cut down the entry barrier cost for running the
%business. Another important point is that the implementation of open standards
%are well verified and they are widely acceptable.}
Many organizations are involved in various standardization efforts on the common
theme of clouds. Notable among them are the working groups operating within the
Open Grid Forum (OGF)~\cite{OGF} umbrella. Other prominent industry consortiums
active in cloud standardization effort are Distributed Management Task Force,
Inc. (DMTF)~\cite{DMTF}, and the Storage Networking Industry Association
(SNIA)~\cite{SNIA}. In this section we will summarize key open cloud standards
that have emerged and point out the cloud component they try to standardize.

The following open standards do help build bridges towards the goal of achieving
user applications and cloud providers interoperability. A significant progress
has been made for pivotal elements such as storage, infrastructure management,
and application description formats, but there still remains much work to be
done to reach the final destination.

\subsection{OGF OCCI}
Open Cloud Computing Interface (OCCI)~\cite{OCCI-1}~\cite{OCCI-2}~\cite{OCCI-3}
proposed standard from the OGF OCCI-Working Group (WG) attempts to standardize
the RESTful protocol and API for management tasks. Initially it was intended
for management of IaaS clouds including deployment, autonomic scaling, and
monitoring, but the standard is quite extensible and can be used for PaaS and
SaaS services. The standard is made of three categories namely \textit{OCCI
Core}, \textit{OCCI Renderings}, and \textit{OCCI Extensions}. In the current
release (version 1.1), it supports HTTP rendering and provides infrastructure
extensions to deal with IaaS clouds.

\subsection{OGF WS-Agreement}
Web Services Agreement Specification (WS-Agreement)~\cite{WS-Agreement} is the
standard specification for web services protocol needed for service level
agreement between the two parties, i.e., the customer and the provider.
This specification uses XML for specifying
the agreement and the agreement templates. It consists of three composable parts
that describe agreement, schema for describing an agreement template, and
operation for managing the lifecycle of the service including monitoring of
agreement states.

\subsection{DMTF CIMI}
Cloud Infrastructure Management Interface (CIMI)~\cite{CIMI} is a
work-in-progress standardization effort within the DMTF consortium that targets
management of resources within the IaaS domain. It implements a REST
interface over HTTP and defines the REST APIs for both XML as well as JSON
rendering. CIMI attempts to provide first-class support to Open Virtualization
Format standard. This work attempts to provide a RESTful management interface
for common IaaS components including machines, networks, volumes, etc.

\subsection{DMTF OVF}
OVF stands for Open Virtualization Format~\cite{OVF-1} and aims to completely
describe a virtual appliance comprised of any number of virtual machines in a
standard and portable format. DMTF advertises this format as vendor-neutral as
it contains no reference to any current vendor-specific information. Written as
an XML file, it features descriptions of most of the components of such an
appliance:
\begin{itemize}
 \item VMs' hardware (CPU, Memory...) and contextualisation
informations;
 \item Disks and images used;
 \item Networking;
 \item Startup order of the different VMs.
\end{itemize}
This format is portable, being platform neutral, and is extensible by the
end-users if needed. DMTF is working towards the next version of this standard
with better support for VM contextualization~\cite{OVF-2}.

\subsection{SNIA CDMI}
Cloud Data Management Interface (CDMI)~\cite{CDMI} defines a RESTful interface
that allows cloud applications and users to retrieve and perform operations on
the data from the cloud. The interface allows capability discovery of storage
elements of the cloud. It also allows administrators to manage the containers,
i.e., metadata, and user accounts and credentials pertaining to the cloud
storage.

\section{Contrail: Striving Towards Interoperability}
European project Contrail~\cite{Contrail-1}~\cite{Contrail-2} is developing a
complete cloud platform which integrates a feature-rich PaaS offering on top of
a federated IaaS cloud providers. An end user of Contrail Cloud Federation
(CCF) will have the ability to do live migration of her applications from one
provider to another. The CCF will have an extensive SLA support along with
required monitoring mechanisms to enforce and manage the negotiated SLAs
between the customers and the providers. The CCF will incorporate an extensive
set of dedicated security suites to manage the authentication, authorization, VM
isolation, and other security needs of the federation and end users. The
CCF is being developed as an open source project with periodic public releases
of the software suite~\cite{Contrail-2}.

The CCF supports DMTF's OVF standard without introducing non-standard
extensions. The project currently supports OpenNebula IaaS clouds but plans to
provide support for OpenStack clouds along with commercial public cloud
providers such as Amazon EC2. Thus CCF facilitates cloud application portability
between providers by translating the standard OVF descriptor into native VM
templates as understood by the various supported IaaS clouds. The CCF comes
with its own virtual infrastructure network (VIN)~\cite{VIN} module that allows
cloud applications to be deployed across multiple providers in a split manner
and still maintaining secure communication channels between different VMs in the
application through IPSec tunnels. If the user's VM needs to be deployed over a
public (non-Contrail) cloud, the VM is prepared with a VIN agent inside to
enable safe networking with the rest of the VMs inside the Contrail federation.

The CCF authentication modules are incorporating several widely use protocol
such as OAuth and Shibboleth~\cite{Shibboleth} and the attributes repository is
being designed to be easily extensible in order to provide support for easy
incorporation of 3rd party attribute repositories so as to ease
migration of user accounts and attributes into CCF.

\begin{figure}[!ht]
  \centering
  \includegraphics[scale=0.4]{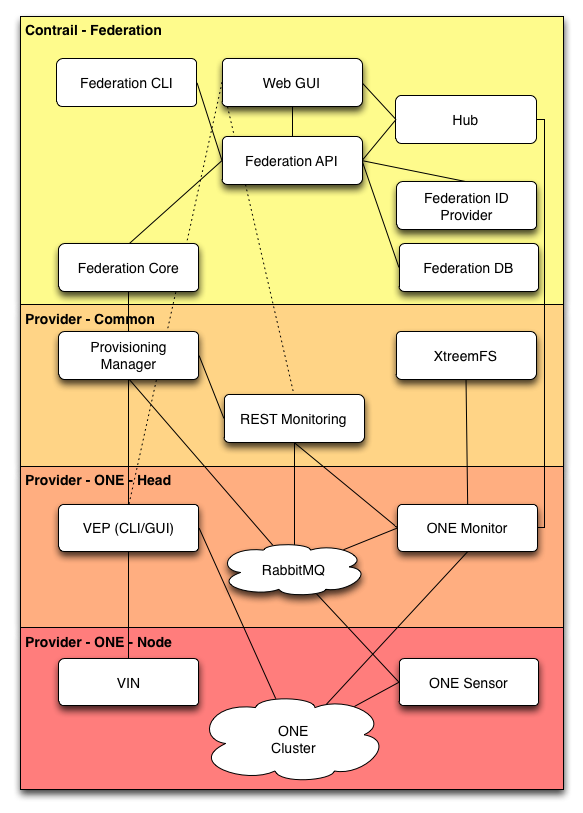}
  \caption{Module Level View of the Contrail Software Architecture (source:
Contrail Release 1.0 Administrator Guide)}
  \label{fig:contrail-module-arch}
\end{figure}

Figure \ref{fig:contrail-module-arch} shows the classification of CCF
software suites released as public release 1.0 into four major categories
namely \textit{federation}, \textit{provider-common},
\textit{provider-ONE-head} and \textit{provider-ONE-node} category. The security
components will be fully integrated with the rest of the modules in subsequent
releases. 

\subsection{Virtual Execution Platform}
Virtual Execution Platform software is installed at the cloud service provider
end and it enables the participation of the cloud in the CCF. It does proper
VM contextualization and OVF application lifecycle management. Additionally it
provides application metrics periodically to the federation modules to help
with SLA monitoring and enforcement. VEP component is being developed so as to
enable non-contrail cloud application developers use VEP in their software
roadmap. In order to be interoperable, a REST interface based on the upcoming
DMTF's CIMI~\cite{CIMI} standard is being designed and developed additionally to
the native REST interface VEP already exposes for integration with rest
of the CCF modules.

\begin{figure}[!ht]
  \centering
  \includegraphics[scale=0.34]{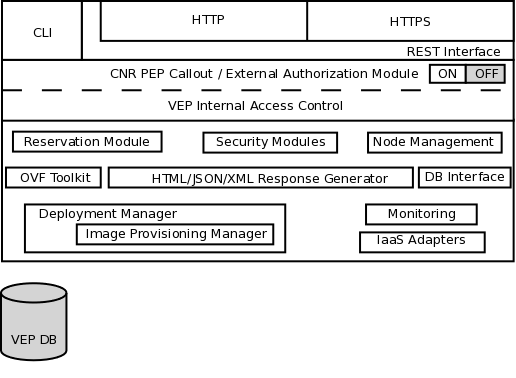}
  \caption{VEP Architecture}
  \label{fig:contrail-vep-arch}
\end{figure}

Figure \ref{fig:contrail-vep-arch} shows the VEP architecture. The OVF toolkit
is the module responsible for performing OVF validation, parsing, and template
generation for the target cloud.

\begin{figure}[!ht]
  \centering
  \includegraphics[scale=0.24]{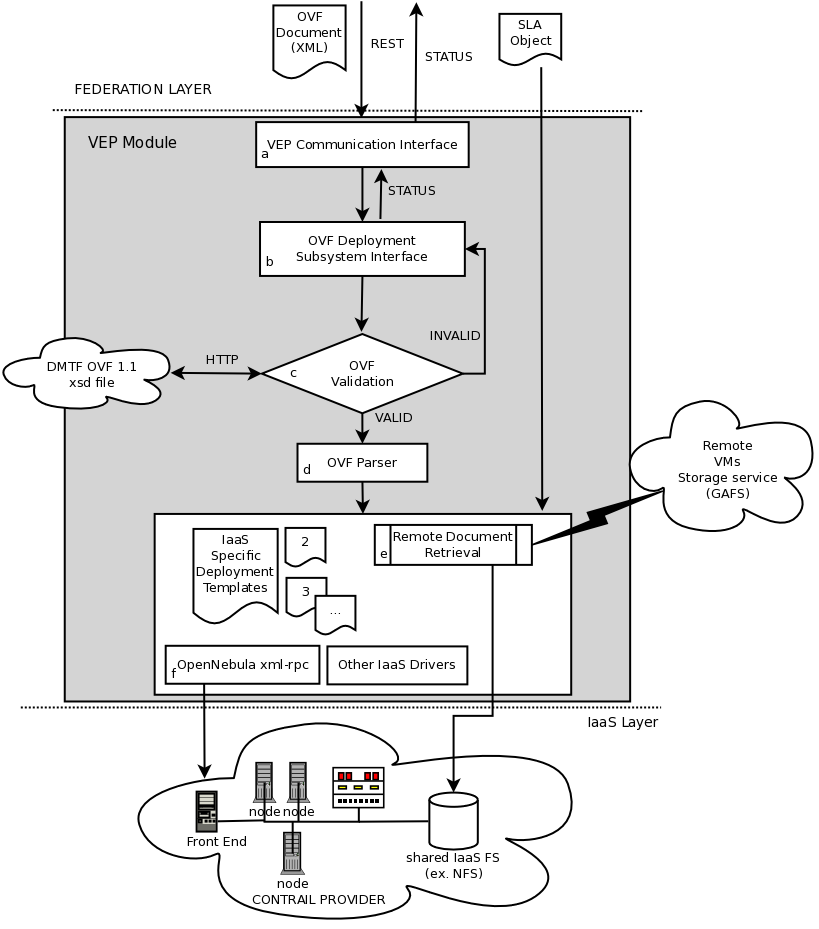}
  \caption{OVF Centric VEP Deployment Taskflow}
  \label{fig:contrail-vep-ovf}
\end{figure}

Figure \ref{fig:contrail-vep-ovf} shows the OVF centric VEP application
deployment task-flow. Contrail and VEP supports only standard OVF descriptions
without any extensions. When an application is registered at VEP, a schema
validation is done before acccepting the application for deployment. Depending
on the IaaS cloud at the provider, compatible VM templates are generated before
the application is deployed on the cloud. If necessary, the VM images could be
retrieved from remote VM datastores into the local provider's filesystem.

VEP implements a two-tier access control on incoming REST requests, one using
internal access-control rules, and an optional entry authorization check using
an external authorization module which can be the one provided by Contrail or 
any compatible one providing the same kind of service. In the subsequent VEP releases, a certificate
delegation module will be incorporated that will allow for a better VM security
by including time-limited role-specific deletegated X.509 certificates to be
passed in the VMs during contextualization phase.

In the above couple of paragraphs we have highlighted key modules and features
of the CCF that helps improve interoperability and portability. Description of
all the modules is beyond the scope of this paper. An interested reader is
encouraged to read the published deliverables and technical
reports~\cite{Contrail-D2.1}~\cite{Contrail-D2.2}~\cite{Contrail-D5.1}~\cite{
Contrail-D5.2}~\cite{Contrail-D7.1}~\cite{Contrail-D7.3} from respective
consortium partners.

Contrail project is embracing open standards namely DMTF's OVF specification
and DMTF's upcoming CIMI specification for enabling independently developed
cloud services to interact with the federation services. An end user will have
the capability of her checkpointed application to be exported as an OVF thereby
allowing her to migrate her application to any provider that supports OVF.

The provision of supporting multiple authentication standards including OAuth
standard, OpenID and Shibboleth and further using an attribute server that can
be updated with 3rd party attributes will allow for an easy integration /
migration of other cloud services with Contrail.

Contrail project will include OVF translation modules for several cloud
technologies, including OpenNebula, OpenStatck and public clouds. This will
enable deployment of OVF applications over Contrail supported cloud technologies
even if the such providers themselves may not support OVF standard. At the
moment Contrail only supports OpenNebula 2.2.1 and OpenNebula 3.4.1 clouds.

\section{Interoperability: Missing Pieces}
While using platform independent application description formats such as OVF,
and a standard cloud management API such as OCCI and application management API
such as CIMI have contributed to improvements in the interoperability scene,
there is a lot more to be done before we achieve seamless interoperability and
portability of cloud resources and end user applications.

\subsection{Credentials Standardization}
Security is a big concern in cloud computing. Various approaches are being
undertaken to limit the exposure of virtual application to threats. Securing
network communication using virtual LANs (vLAN), SSH tunnels, X.509 certificate
based user access, etc. are a few measures being adopted towards this goal.
Still there is a lack of procedural standards in this field.

User attributes at one provider can not be easily transferred to another
provider because of lack of standards for attributes based access control.

\subsection{Network Standardization}
Each cloud suite uses their internal network element representation making it
difficult to port an application descriptor tailored for one cloud to another
completely different cloud. It is agreed that each application has specific
networking needs, but for a case where a simple inter-VM communication is
desired, there should a standard way of describing such a requirement that
would work across all cloud providers. Standardization of common use cases for
virtual networks would help a lot in achieving true interoperability among
cloud and portability to the end user applications.

\section{Conclusion}
In this paper we have tried to identify some of the challenges facing cloud
computing. We have surveyed the ongoing standardization efforts for management
of cloud services and infrastructure. We have presented a brief description
about the European software project Contrail, and how it aims to improve
portability of cloud applications and achieve interoperability with other cloud
services and management tools. We have also provided a few challenges that
should be addressed if any practical interoperability of services and
portability of end users' cloud applications are to be achieved.


\section*{Acknowledgment}
Contrail is a European Commission funded project under FP7 program and is funded
by grant 257438. The authors would like to acknowledge the work done by all
Contrail Consortium members and teams.



\bibliographystyle{IEEEtran}
\bibliography{IEEEabrv,paper}
%\begin{thebibliography}{1}


%\end{thebibliography}

\end{document}